\documentclass[twocolumn, superscriptaddress, secnumarabic, amssymb, showpacs, nobibnotes, aps, prb]{revtex4-2}
\usepackage{graphicx}
\usepackage[font=footnotesize]{subcaption}
\usepackage{wrapfig}
\usepackage{dcolumn}
\usepackage{bm}
\usepackage{color}
\usepackage{tabularx}
\usepackage{csquotes}
\usepackage[normalem]{ulem}
\usepackage{xcolor}
\usepackage{siunitx}[locale=US]
\DeclareSIUnit\angstrom{\text{Å}}
\usepackage{upgreek}
\usepackage{physics} 

\newcommand{\LOFOS}{La\textsubscript{2}O\textsubscript{2}Fe\textsubscript{2}OSe\textsubscript{2}}
\newcommand{\LOFMxOS}{La\textsubscript{2}O\textsubscript{2}Fe\textsubscript{2-x}
Mn\textsubscript{x}OSe\textsubscript{2}}
\newcommand{\LOMOS}{La\textsubscript{2}O\textsubscript{2}Mn\textsubscript{2}OSe\textsubscript{2}}
\newcommand{\symmetry}{\textit{I}4/\textit{mmm}}

\usepackage{url}
\begin{document}

\title{Confirmation of complex non-collinear Fe(II) spin ordering in the iron oxychalcogenide \LOFOS\, by broad range $^{139}$La-NQR}

\author{Tillmann Weinhold}\affiliation{Institute of Solid State and Materials Physics, IFMP, TUD Dresden University of Technology, Dresden, Germany}

\author{Jerin Abraham George}\affiliation{Leibniz Institute for Solid State and Materials Research Dresden, Helmholtzstr. 20,
D-01069 Dresden, Germany}

\author{Cornelius Herrmann}\affiliation{Institute of Solid State and Materials Physics, IFMP, TUD Dresden University of Technology, Dresden, Germany}

\author{Domenic Nowak}\affiliation{Leibniz Institute for Solid State and Materials Research Dresden, Helmholtzstr. 20,
D-01069 Dresden, Germany}

\author{Rajib Sarkar}\affiliation{Institute of Solid State and Materials Physics, IFMP, TUD Dresden University of Technology, Dresden, Germany}

\author{Felix Seewald}\affiliation{Institute of Solid State and Materials Physics, IFMP, TUD Dresden University of Technology, Dresden, Germany}

\author{Rowena Wachtel}\affiliation{Leibniz Institute for Solid State and Materials Research Dresden, Helmholtzstr. 20,
D-01069 Dresden, Germany}

\author{Sabine Wurmehl}\affiliation{Leibniz Institute for Solid State and Materials Research Dresden, Helmholtzstr. 20,
D-01069 Dresden, Germany}

\author{Bernd B\"uchner}\affiliation{Leibniz Institute for Solid State and Materials Research Dresden, Helmholtzstr. 20,
D-01069 Dresden, Germany}

\author{Hans-Henning Klauss}\affiliation{Institute of Solid State and Materials Physics, IFMP, TUD Dresden University of Technology, Dresden, Germany}

\date{21. October 2026}

\begin{abstract}
    The transition metal oxychalcogenide \LOFOS \, exhibits a non-collinear spin-ice like antiferromagnetic ordering below $T_{\mathrm N} \approx \SI{90}{\kelvin}$ due to complex magnetic exchange interactions and axial single-ion anisotropy of Fe$^{2+}$ spins. The magnetic structure was deduced/determined from combined $^{57}$Fe Mössbauer spectroscopy and $^{139}$La-NMR on polycrystalline specimen. In this work, we present zero-field $^{139}$La-NQR investigations on polycrystalline \LOFOS \, in a broad frequency and temperature range between $\num{2.8} - \SI{20}{\mega\hertz}$ and $\num{4.2} - \SI{100}{\kelvin}$. The measured  resonances at $\nu_Q$, $2 \nu_Q$, and $3 \nu_Q$ are fully consistent with the previous NMR studies and also with the magnetic structure.
    A substantial modification of the electric field gradient at the La-position below  $T_{\mathrm N}$ is inconsistent with the current investigations. 
\end{abstract}

\maketitle

\section{Introduction}
The physics of altermagnets forms one of the most vibrant areas in solid state magnetism research in recent years \cite{doi:10.1126/sciadv.aaz8809, doi:10.7566/JPSJ.88.123702, PhysRevX.12.031042, PhysRevX.12.040501, GUO2023100991}.
Altermagnets offer a new perspective for Terahertz spintronics applications of compensated antiferromagnets, since they exhibit electronic band structures with large spin-splitting in the \si{\eV} range in a zero net magnetization system. The spin-split character of the electron bands relies on a non-relativistic effect, namely an anisotropic electron hopping within the two spin sublattices \cite{doi:10.1021/jacs.4c14503}, which is caused by the combination of time-reversal-symmetry breaking and spatial rotation between the spin sublattices. Tetragonal systems with the in-plane Lieb-lattice structural motive like KV\textsubscript{2}Se\textsubscript{2}O are inherently proposed candidates for altermagnetic properties with d-wave symmetry \cite{Jiang2025}. D-wave altermagnets are particularly interesting for applications due to their large 90$^{\circ}$ anisotropy of the spin-split electronic band structure.

The transition metal oxychalcogenides \LOFMxOS\ have \symmetry\ symmetry and exhibit the Lieb-lattice structural motive. \LOMOS, showing collinear antiferromagnetic order with Mn$^{2+}$ spins aligned along the c-axis, is proposed as a correlated insulating d-wave altermagnet \cite{PhysRevMaterials.9.024402}.

\begin{figure}[!htb]
    \centering
    \begin{subfigure}{0.25\columnwidth}
        \centering
        \includegraphics[width=\textwidth]{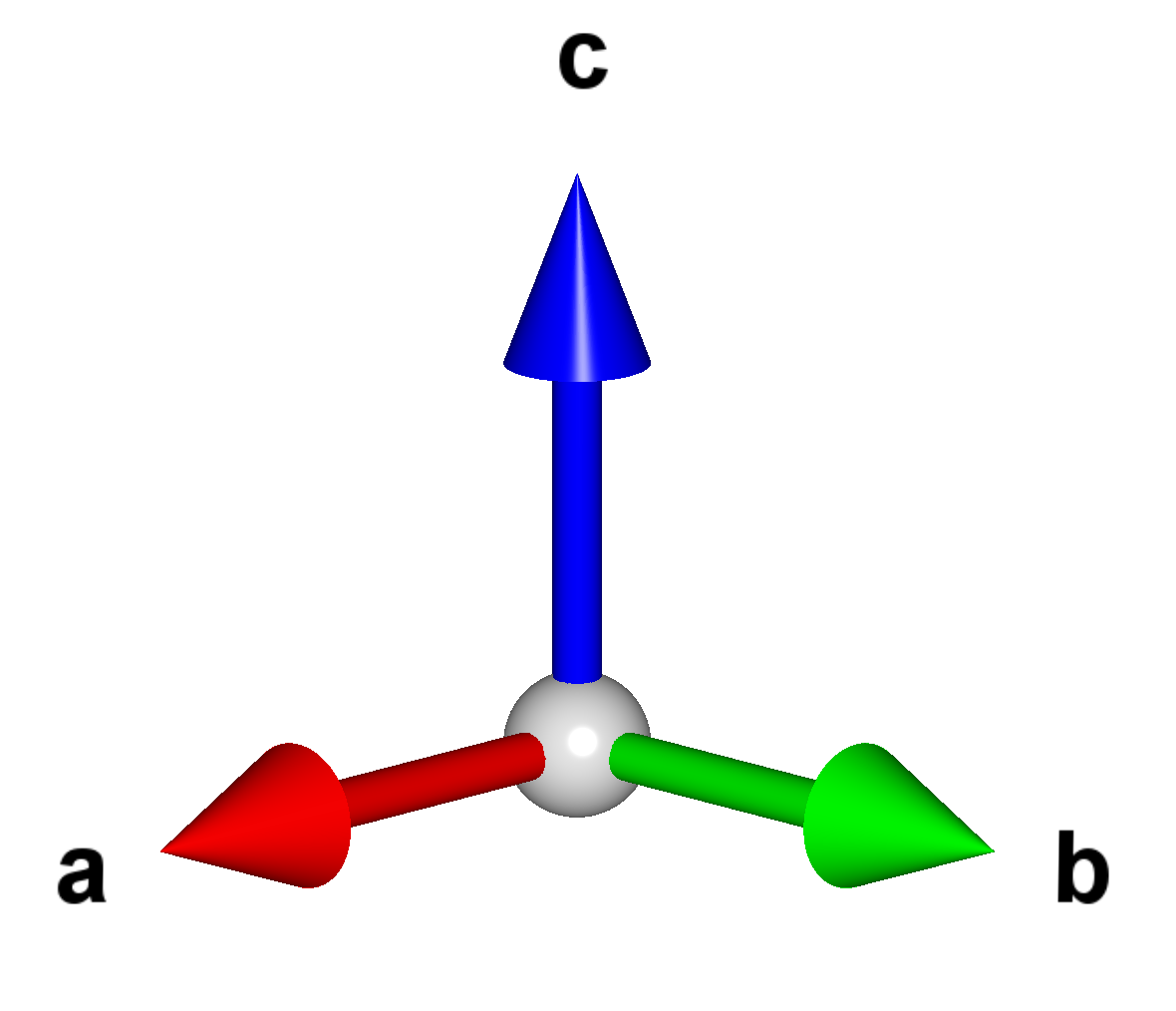}
    \end{subfigure}
    ~
    \begin{subfigure}{0.4\columnwidth}
        \centering
        \includegraphics[width=\textwidth]{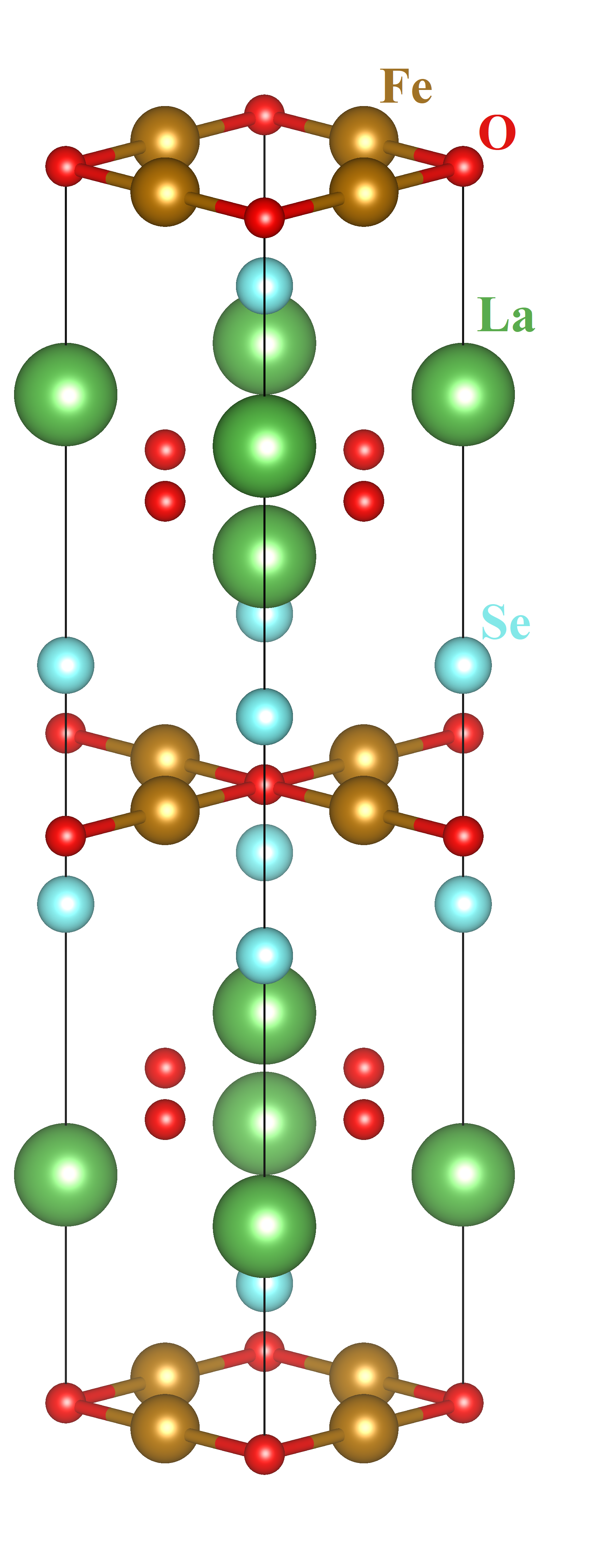}
        \vspace{-2.5\baselineskip}
    \end{subfigure}
    \caption{\label{fig:cyrstal_structure}Unit cell of \LOFOS. Brown spheres indicate the Fe positions, red spheres the O position. Green and light blue spheres indicate La and Se position, respectively. Figure created with VESTA3 \cite{Vesta3}.}
\end{figure}

\begin{figure}[!htb]
\centering
\includegraphics[width=0.98\columnwidth]{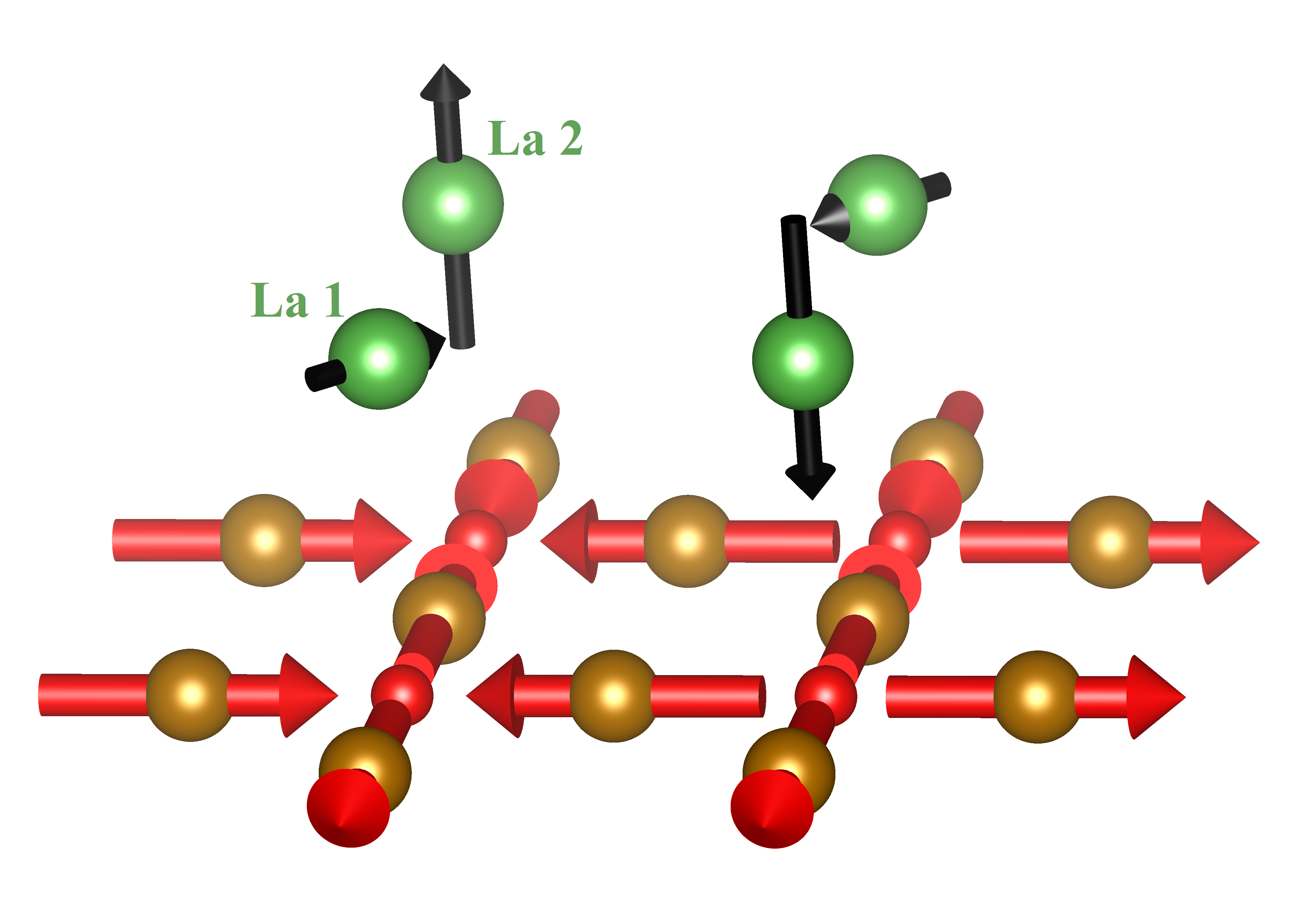}
\caption{ Magnetic structure of \LOFOS\ as determined by Günther {\em et al.} \cite{PhysRevB.90.184408} and McCabe {\em et al.} \cite{PhysRevB.89.100402}.
The atoms are colored as in Figure \ref{fig:cyrstal_structure}. Red arrows indicate the direction of iron moments below $T_\mathrm{N}$. Black arrows at the La sites indicate the direction of the dipolar magnetic field $\vec{B}$ at the La positions induced by the iron moments. Two different La sites are present in the ordered phase due to different magnetic field strength and direction. These La sites are labeled as La$\,$1 and La$\,$2, consistent with the labeling by Günther {\em et al.} \cite{PhysRevB.90.184408} and Hussain {\em et al.} \cite{PhysRevB.103.L081105}.
}
\label{fig:magnetic_order}
\end{figure}

The iron system \LOFOS\  is an insulator \cite{PhysRevB.90.184408}. 
La\textsubscript{2}O\textsubscript{2}\textsuperscript{2+} layers separate the Fe\textsubscript{2}OSe\textsubscript{2}\textsuperscript{2-} layers stacked along the c-axis, as shown in Fig.~\ref{fig:cyrstal_structure}. 
Below the N\'eel temperature $T_\mathrm{N}$\,=\,\SI{90}{\kelvin} the system  develops antiferromagnetic order \cite{PhysRevB.82.214419, PhysRevB.81.214433, PhysRevB.86.125122}. 
Fe\textsuperscript{2+} with an atomic 3d\textsuperscript{6} $S\,=\,2, L\,=\,2$ high spin ground state and considerable spin-orbit coupling exhibits a strong axial magnetocrystalline anisotropy in the antiferromagnetically ordered state. 
In contrast to earlier neutron diffraction studies \cite{PhysRevB.81.214433, Landsgesell_2013}, from which a collinear AFM order below \SI{90}{\kelvin} was concluded, a combined local probe study by $^{57}$Fe Mössbauer spectroscopy, $^{139}$La-NMR and $\mu$SR in 2014 proved that the Fe spins align along the O - Fe - O bonds with 90 degree in-plane non-collinear magnetic order on the Lieb lattice \cite{PhysRevB.90.184408}. 
In that study an in-plane Fe order symmetry similar to Sr\textsubscript{2}F\textsubscript{2}Fe\textsubscript{2}OS\textsubscript{2} \cite{doi:10.1021/ja711139g} was deduced from the appearance of two  different $^{139}$La-NMR subspectra with equal intensity yet different internal magnetic field strength and orientation in the ordered state. 
A later neutron diffraction study by McCabe {\em et al.} \cite{PhysRevB.89.100402} supports the non-collinear magnetic order.
Figure \ref{fig:magnetic_order} displays the non-collinear order as deduced by Günther {\em et al.} \cite{PhysRevB.90.184408} and McCabe {\em et al.} \cite{PhysRevB.89.100402}.
Recently, \LOFOS\ attracted further interest due to the report of short-range orthorhombicity both in the paramagnetic phase and in the magnetic phase based on the analysis of the neutron pair distribution function \cite{Freelon2021}. 

In 2021 Hussain {\em et al.} reported a zero external magnetic field  $^{139}$La-NQR study on \LOFOS\ \cite{PhysRevB.103.L081105}. 
The $^{139}$La NMR spectra for polycrystalline \LOFOS \, show rather broad resonance lines due to the continuous distribution of relative angles formed by the external magnetic field and the principle axes of the local electric field gradient \cite{PhysRevB.90.184408}. Those broad lines collapse to discrete lines in the $^{139}$La NQR spectra.
Below $T_{\mathrm N}$ the discrete NQR resonances are expected to shift due to the internal transferred magnetic hyperfine field. 
 In contrast to this expectation, Hussain {\em et al.} reported a discontinuous {\emph splitting} of the resonance signals at the $2 \, \nu_Q$ and $3 \, \nu_Q$ transitions into two subspectra below $T_\mathrm{N}$.
  For one, referred to La$\,$1, the frequency is constant from above $T_{\mathrm N}$  down to lowest temperature. The frequency of the second subspectrum, assigned as La$\,$2, rises abruptly at $T_{\mathrm N}$  and shows a strong temperature dependence. At 60 K, the La$\,$2 signal crosses the resonance line of La$\,$1 that is fixed in frequency. Below 30 K, also the La$\,$2 spectrum  saturates and becomes temperature independent.
The authors interpreted these findings as a magnetically induced local lattice anomaly leading to two  La-sites with different electric field gradients below $T_\mathrm{N}$. 
In 2023, Yang {\em et al.} reported a hydrostatic pressure induced transition from a Mott insulator to a ferromagnetic Weyl metal in \LOFOS \, at $\approx  35 $GPa \cite{Yang2023}.

In this study, we report simulations of zero external magnetic field NQR/NMR spectra for all possible transitions of the $^{139}$La (I=7/2) nucleus further supported by additional NQR measurements in a wide frequency range between 2.8 and 20 MHz in the temperature range 4.2 to 100 K. 
We show that neither a splitting nor a strong
change of the EFG at the La-sites below $T_{\mathrm N}$ is needed to
explain the experimental observations. Instead, all measured NQR spectra, including those from ref. \cite{PhysRevB.103.L081105}, are well modelled assuming two La-sites with different transferred
magnetic hyperfine fields generated by the non-collinear
magnetic ordering of the Fe-spins.

\section{Synthesis}
Polycrystalline samples of \LOFOS\ were prepared by a solid-state reaction using La\textsubscript{2}O\textsubscript{3}, Fe, and Se powders, following a similar procedure to that reported by Günther {\em et al.} \cite{PhysRevB.90.184408}. 
Initially, the La\textsubscript{2}O\textsubscript{3} powder was dried at \SI{1000}{\degreeCelsius} for \SI{12}{\hour} to remove moisture. All precursors were weighed in the stoichiometric ratio, ground in an agate mortar, and pressed into \SI{5}{\milli\meter}-diameter pellets in an Ar-filled glovebox (MBraun, Germany). The pellets were sealed in a \SI{15}{\milli\meter}-diameter quartz tube under vacuum and heated in a GERO high-temperature furnace with the following sequence: ramp to \SI{600}{\degreeCelsius} at \SI{2}{\degreeCelsius\per\minute} and dwell for \SI{12}{\hour}, ramp to \SI{800}{\degreeCelsius} at \SI{1}{\degreeCelsius\per\minute} and dwell for \SI{1}{\hour}, ramp to \SI{1100}{\degreeCelsius} at \SI{2}{\degreeCelsius\per\minute} and dwell for \SI{96}{\hour}. After sintering, the furnace was cooled to room temperature.  To improve homogeneity, the resulting pellets were reground and subjected to a second sintering step at \SI{1100}{\degreeCelsius} for \SI{72}{\hour}. 
Structure and phase purity were checked by powder X-ray diffraction and scanning electron microscopy along with energy-dispersive X-ray spectroscopy.

\section{Experiment}
\begin{figure*}[!htb]
    \centering
    \includegraphics[width=0.95\textwidth]{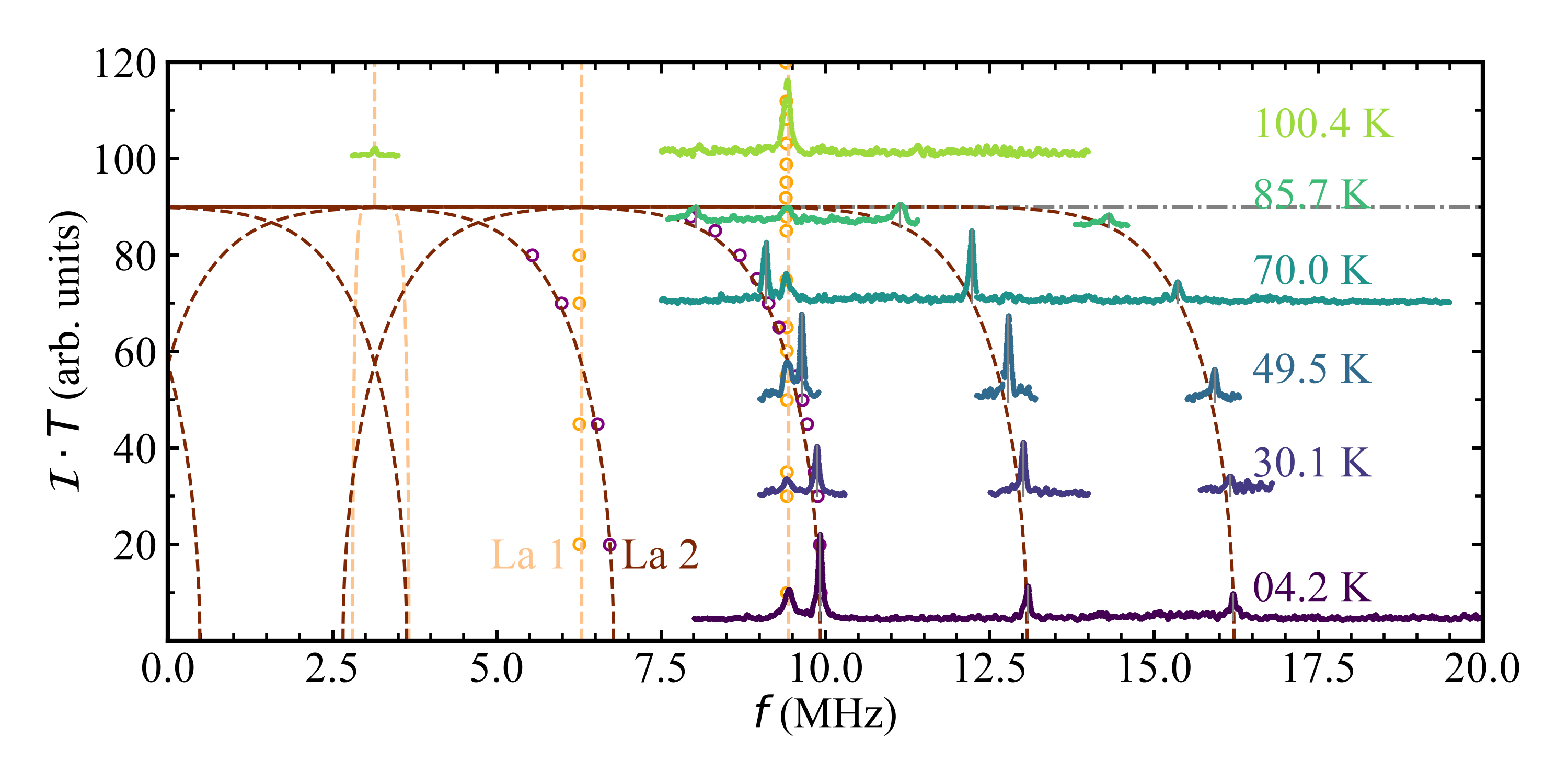}
    \caption{\label{fig:spectra_and_theory}\textsuperscript{139}La NQR spectra of \LOFOS\ at different temperatures. The base line of each spectrum is shifted by a factor that reflects the evolution of temperature. 
    The dashed gray horizontal line marks the  ordering temperature of $T_\mathrm{N} \approx \SI{90}{\kelvin}$. The open circles mark the peak positions of the NQR signals found by Hussain {\em et al.} \cite{PhysRevB.103.L081105}. The dashed bright orange and dark orange lines indicate the expected peak positions according to the model described in section IV. and introduced by Günther {\em et al.} \cite{PhysRevB.90.184408} for site La$\,$1 and La$\,$2, respectively.}
\end{figure*}
Pulsed \textsuperscript{139}La nuclear quadrupole resonance (NQR) measurements have been performed on \LOFOS\ using a \emph{LapNMR Pulse Generator}, an \emph{AMT M3205 Pulse Amplifier}, and the \emph{Tecmag TNMR} software.
The NQR frequency has been varied from \SIrange{7.5}{20}{\mega\hertz} in \SI{0.1}{\mega\hertz} steps for different temperatures from \SIrange{4.2}{100.4}{\kelvin}.
Additionally, measurements at frequencies between \SI{2.8}{\mega\hertz} and \SI{3.5}{\mega\hertz} in steps of \SI{0.05}{\mega\hertz} have been performed at \SI{100.4}{\kelvin}. 
No signal has been observed in that frequency regime below $T_\mathrm{N}$.
Each time spectrum has been multiplied by a gaussian envelope function with a width $\sigma = \SI{8}{\micro\second}$ centered at the position of the echo (Gaussian apodization) to suppress detected noise appearing over the whole time frame.
The time spectra were then Fourier transformed to obtain the frequency dependent intensities. only the frequencies of the used excitation frequency $\pm\SI{100}{\kilo\hertz}$ have been taken into account.

The resulting spectra are shown in Figure \ref{fig:spectra_and_theory}.
The dashed lines correspond to the theoretical predicted resonances (see section \ref{sec:theory}).
Above the ordering temperature of \SI{90}{\kelvin} only one resonance line \SI{9.4}{\mega\hertz} could be identified  between \SI{7.5}{\mega\hertz} and \SI{14}{\mega\hertz}.
This is the \textsuperscript{139}La $3 \,\nu_\mathrm{Q}$ signal as been identified by Hussain {\em et al.} \cite{PhysRevB.103.L081105}.
Below the ordering temperature three additional resonance peaks appear in the investigated frequency regime. These lines are equidistantly separated by $\nu_\mathrm{Q}$ at all measured temperatures but  exhibit an order parameter like shift as a function of temperature.
In contrast, the frequency of the original peak, which is considerably reduced in intensity, is temperature independent within the frequency resolution.

The separation of the three additional resonance peaks by exactly $\nu_\mathrm{Q}$ and their order parameter like homogeneous shift as a function of temperature clearly confirm the presence of a magnetic field at the La position.
Not all seven expected resonance peaks have been observed due to frequency limitations of the LC circuit.  The strength of the magnetic saturation field at $T\approx\SI{0}{\kelvin}$ was determined by the absence of resonance peaks above \SI{16.3}{\mega\hertz} to \SI{1.127(10)}{\tesla}.
A second La position, that does not observe any magnetic field of comparable strength, is needed to explain the temperature independent resonance peak.
Magnetically induced local lattice anomalies (suggested by Hussain {\em et al.} \cite{PhysRevB.103.L081105}) are not sufficient to explain the  appearance of additional resonance peaks, which are split exactly by $\nu_\mathrm{Q}$.

\section{Simulations}
\label{sec:theory}
\begin{figure*}[!htb]
    \centering
    \begin{subfigure}{0.48\textwidth}
    \centering
        \includegraphics[width=\textwidth]{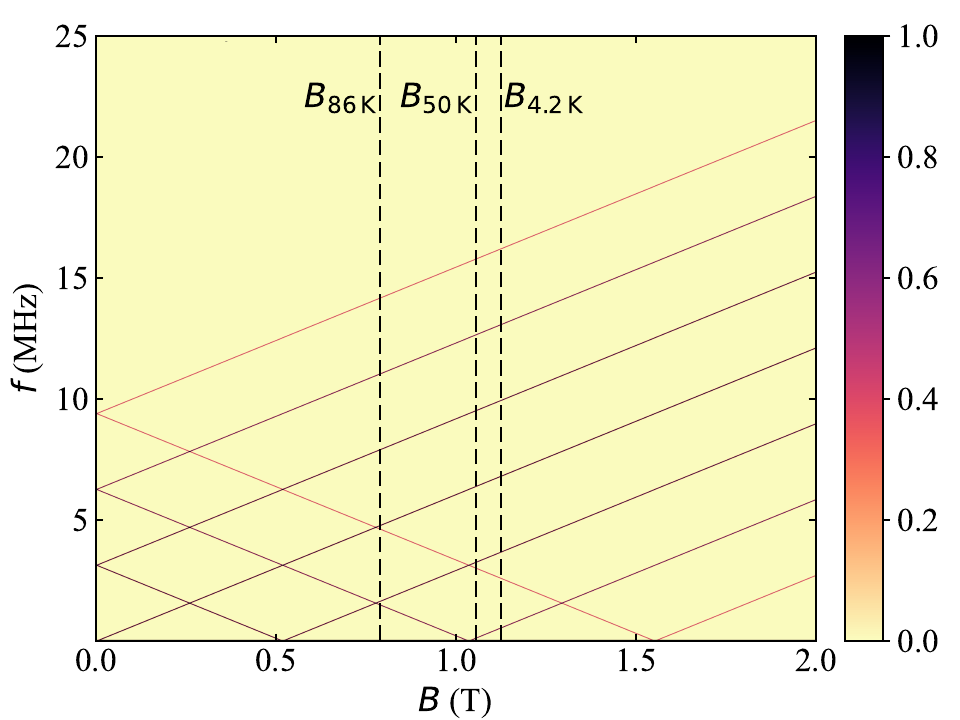}
        \caption{\label{fig:SIM_B_par_V}$\vec{B}\parallel \vec{V}_\mathrm{zz}$} 
    \end{subfigure}
    ~
    \begin{subfigure}{0.48\textwidth}
        \centering
        \includegraphics[width=\textwidth]{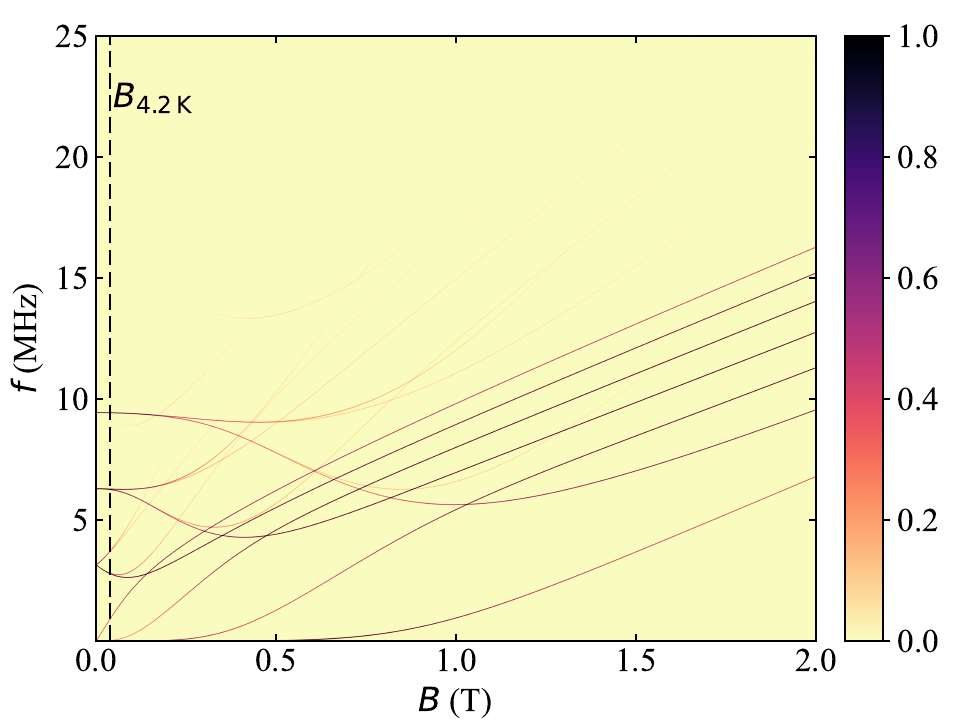}
        \caption{\label{fig:SIM_B_perp_V}$\vec{B}\perp \vec{V}_\mathrm{zz}$}
    \end{subfigure}
    \caption{\label{fig:SIM_B_V}Simulation of the NQR resonance positions for the case of a magnetic field being (a) parallel to the main component of the EFG corresponding to site La$\,$2 in Figure \ref{fig:magnetic_order} and (b) perpendicular to the main component of the EFG corresponding to site La$\,$1 in Figure \ref{fig:magnetic_order}. Parameters used are $\nu_\mathrm{Q}=\SI{3.147}{\mega\hertz}$ and $\gamma_\mathrm{^{139}La} = \SI{6.06}{\mega\hertz\per\tesla}$. The relative intensities of the various resonance peaks (color-coded as in the vertical bars) have been approximated by calculating the peak intensities for \num{2000} random orientations of the rf excitation pulse. Dashed black lines indicate the strength of magnetic field expected at the specified temperature.}
\end{figure*}
According to Günther {\em et al.} \cite{PhysRevB.90.184408} two different La sites  exist in the magnetically ordered state, one, La$\,$2, observing a larger magnetic field along the $c$ axis and thus parallel to the main component of the EFG $V_\mathrm{zz}$ and the other, La$\,$1, observing a small magnetic field in the $a-b$ plane (crf. black arrows in Fig. 2).
To compare the observed spectra with the suggested model, \textsuperscript{139}La NMR spectra have been simulated for both cases $\vec{B}\parallel\vec{V}_\mathrm{zz}$ and $\vec{B}\perp\vec{V}_\mathrm{zz}$ for various strengths of magnetic field $B$ assuming $\nu_\mathrm{Q} = \SI{3.147}{\mega\hertz}$ (see Fig. 4).
The employed Hamiltonian to calculate energies and eigenstates is
\begin{equation}
\begin{aligned}
    \mathcal{H} =& \mathcal{H}_\mathrm{Z} + \mathcal{H}_\mathrm{Q} = \gamma_\mathrm{^{139}La}\hbar B \mathcal{I}_z + \\
    &\frac{eQ}{4\mathcal{I}\pqty{2\mathcal{I}-1}} [V_{zz}\pqty{3\mathcal{I}_z^2-\mathcal{I}^2} \\
    &+ \pqty{V_{zx}+\mathrm{i}V_{zy}}\pqty{\mathcal{I}^-\mathcal{I}_z+\mathcal{I}_z\mathcal{I}^-} \\ 
    &+ \pqty{V_{zx}-\mathrm{i}V_{zy}}\pqty{\mathcal{I}^+\mathcal{I}_z+\mathcal{I}_z\mathcal{I}^+} \\
    &+ \frac{1}{2}\pqty{V_{xx}-V_{yy}+2\mathrm{i}V_{xy}}\pqty{\mathcal{I}^-}^2 \\
    &+ \frac{1}{2}\pqty{V_{xx}-V_{yy}-2\mathrm{i}V_{xy}}\pqty{\mathcal{I}^+}^2].
\end{aligned}
\end{equation}
Relative intensities are then calculated by treating the excitation pulse of the NMR experiment as a small perturbation and calculating $\bra{f}\cos{\varphi}\sin{\vartheta}\mathcal{I}_x + \sin{\varphi}\sin{\vartheta}\mathcal{I}_y + \cos{\vartheta}\mathcal{I}_z\ket{i}$, where $\ket{i}$ and $\ket{f}$ are eigenstates of $\mathcal{H}$.
To simulate powder spectra, intensities have been calculated \num{2000} times for random directions of the excitation pulse to account for the random orientation of the grains of material. 
The results are shown in Figure \ref{fig:SIM_B_V}.

The obtained frequency and field relations have been used to check the validity of the model described in Günther {\em et al.} \cite{PhysRevB.90.184408}.
According to the model one La site is observing a large magnetic field (about \SI{1}{\tesla} at \SI{50}{\kelvin}) below $T_\mathrm{N}$ mainly parallel to the $c$ axis, thus being parallel to the principal axis of the EFG $V_{zz}$.
The direction of $V_{zz}$ being parallel to the $c$ axis is a consequence of the crystals symmetry, as the rotational axis parallel to $c$ through the La atom has 4-fold symmetry. 
This is consistent both with the distance of all three resonance peaks appearing below $T_\mathrm{N}$ as well as their frequency.
A second La site observes a small magnetic field in the $a-b$ plane according to the model.
The expected magnetic field would be smaller by a factor of approximately \num{30} taking only dipole field calculations into account \cite{PhysRevB.90.184408}.

The temperature dependent resonance peak positions according to the model of Günther  {\em et al.} \cite{PhysRevB.90.184408} are shown in Figure \ref{fig:spectra_and_theory}.
The temperature dependence of the  transferred
magnetic hyperfine fields at the La sites has been assumed to be proportional to the magnetic field at the iron position determined by \textsuperscript{57}Fe Mössbauer spectroscopy (see the Supporting Information).
Assuming the resonance peak at  $\num{3}\, \nu_\mathrm{Q}$ = \SI{9.44}{\mega\hertz} to be temperature independent, only two free parameters are remaining: the low temperature saturation field at the high field La site and $\nu_\mathrm{Q}$.

Our simulations show that there is no detectable change of frequency for the $\num{3}\, \nu_\mathrm{Q}$ resonance peak expected below \SI{100}{\milli\tesla} and even for the $\num{2}\, \nu_\mathrm{Q}$ resonance peak any splitting might not be observable for fields below \SI{50}{\milli\tesla}. 
Therefore, from the measurements shown here and the data published in Hussain {\em et al.} \cite{PhysRevB.103.L081105} it can be concluded that the magnetic field is smaller than \SI{50}{\milli\tesla}, which is in agreement with the proposed value of about \SI{35}{\milli\tesla} for $T\approx\SI{0}{\kelvin}$ \cite{PhysRevB.90.184408}.
To investigate the magnetic field strength at that La site, additional measurements at $f\approx\nu_\mathrm{Q}$ below $T_\mathrm{N}$ would be needed (see lower left part of Fig.4 (b)). These require an improved low pass signal filtering.

\section{Summary}

In this study, we report simulations of zero external magnetic field NQR/NMR spectra for all possible transitions of the  $^{139}$La (I=7/2) nucleus and additional NQR measurements in a wide frequency range between 2.8 and 20 MHz in the temperature range 4.2 to 100 K. 
We show that neither a splitting nor a strong
change of the EFG at the La-sites below $T_{\mathrm N}$ is needed to consistently explain the experimental observations. Instead, all measured NQR spectra, including those from ref. \cite{PhysRevB.103.L081105}, are well modelled assuming two La-sites with different transferred
magnetic hyperfine fields generated by the non-collinear
magnetic ordering of the Fe-spins.
The results of the model calculations and the measurements are in very good agreement. We conclude that the magnetic structure in \LOFOS\ is the one shown in Fig. \ref{fig:magnetic_order}  as proposed by Günther {\em et al.} \cite{PhysRevB.90.184408} and McCabe {\em et al.} \cite{PhysRevB.89.100402}.

Combining the different nuclear probe methods NMR and Mössbauer spectroscopy is a powerful tool to separate collinear single-$\vec{q}$ from non-collinear multi-$\vec{q}$ magnetic structures in polycrystalline specimen of \LOFOS\,, since for $^{139}$La and $^{57}$Fe the electric quadrupole interaction is of similar order of magnitude as the magnetic Zeeman interaction at the nuclear sites.

\section*{Acknowledgments} 
This research was supported by the Deutsche Forschungsgemeinschaft (DFG, German Research Foundation) through projects B01 and C02  of the SFB 1143 (project-id 247310070), by the Würzburg-Dresden Cluster of Excellence ct.qmat, EXC 2147 (project ID 455319354), and by individual grant WU595/14-1. 
J.A.G and S.W. are indebtedly thankful to Nico Grässler for his experimental support and very fruitful discussions as well as R. Kluge and C.G.F. Blum for technical support (all IFW Dresden).

\section*{Supporting Information - \textsuperscript{57}Fe Mössbauer}
\begin{figure}[b]
    \centering
    \includegraphics[width=0.98\columnwidth]{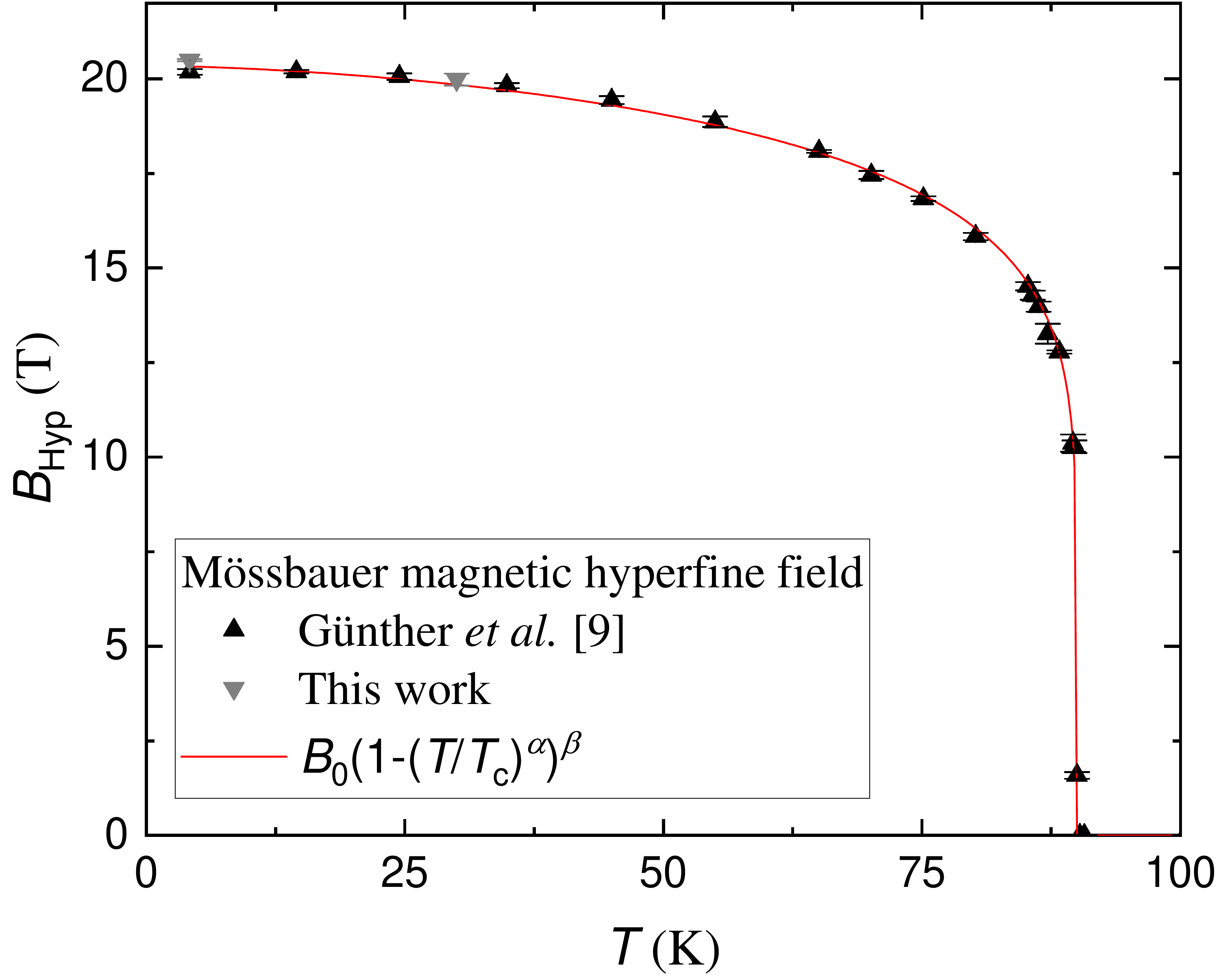}
    \caption{\label{fig:La2O2Fe2OSe2-MBSOrderparameter.pdf}Local magnetic hyperfine field $B_\mathrm{Hyp}$ at the $^{57}$Fe position measured by Mössbauer spectroscopy.
    Black upwards facing triangles are data published by Günther {\em et al.} \cite{PhysRevB.90.184408}. 
    Grey downwards facing triangles are measured on the sample used in this work.
    The red line corresponds to a phenomenological order parameter function (see text).
    }
\end{figure}

Figure \ref{fig:La2O2Fe2OSe2-MBSOrderparameter.pdf} presents the temperature dependent magnetic hyperfine field obtained from \textsuperscript{57}Fe Mössbauer spectroscopy measurements.
The data is described utilizing a phenomenological $B_\mathrm{Hyp} = B_0 \pqty{1-\pqty{\frac{T}{T_\mathrm{c}}}^\alpha}^\beta$ model.
Values of $\alpha = 1.6(4)$, $\beta = 0.134(11) $, $T_\mathrm{c} = $\SI{89.99(5)}{\kelvin} are obtained. The critical exponent $\beta$ =
0.134 is close to that of the magnetic square planar Ising model
($\beta$ = 1/8). The exponent  $\alpha$ describes the behavior for $T \rightarrow 0$ (Bloch law). The value $\alpha= 1.6(4)$ is close to the expectation  $\alpha= 2$ for an antiferromagnet. 

\bibliography{LOFOS}

\end{document}